# Meaningful causal decompositions in health equity research: definition, identification, and estimation through a weighting framework.

John W. Jackson[1]

**Abstract**

Causal decomposition analyses can help build the evidence base for interventions that address health disparities (inequities). They ask how disparities in outcomes may change under hypothetical intervention. Through study design and assumptions, they can rule out alternate explanations such as confounding, selection-bias, and measurement error, thereby identifying potential targets for intervention. Unfortunately, the literature on causal decomposition analysis and related methods have largely ignored equity concerns that actual interventionists would respect, limiting their relevance and practical value. This paper addresses these concerns by explicitly considering what covariates the outcome disparity and hypothetical intervention adjust for (so-called allowable covariates) and the equity value judgements these choices convey, drawing from the bioethics, biostatistics, epidemiology, and health services research literatures. From this discussion, we generalize decomposition estimands and formulae to incorporate allowable covariate sets, to reflect equity choices, while still allowing for adjustment of non-allowable covariates needed to satisfy causal assumptions. For these general formulae, we provide weighting-based estimators based on adaptations of ratio-of-mediator-probability and inverse-odds-ratio weighting. We discuss when these estimators reduce to already used estimators under certain equity value judgements, and a novel adaptation under other judgements.

[1] John W. Jackson, Sc.D., is Assistant Professor in the Departments of Epidemiology and Mental Health at the Johns Hopkins Bloomberg School of Public Health. Correspondence to john.jackson@jhu.edu.

**Introduction**

Health disparities represent differences across socially privileged vs. socially marginalized groups that society considers inequitable, avoidable, and unjust.[1] Interventions that address disparities[2] usually affect risk factors that are overrepresented among marginalized groups. Often their evidence base draws from studies that compare measures of disparities before and after adjustment for a risk factor (the difference method[3]). But the changes seen after such adjustments may be due confounding, selection bias, or information bias. The results may not imply that the risk factor studied is the one to be intervened upon. Causal decomposition methods[4-8] compare to a counterfactual disparity under a hypothetical intervention on a target risk factor. They overcome the limitations of simple adjustment through sound study design and unverifiable assumptions to rule out alternative explanations (bias). They ask a simple question, how disparities in outcomes would change if disparities in a targeted factor (that affects the outcome) were removed.

Unfortunately, estimators used for causal decomposition have ignored how disparities and hypothetical interventions are defined, limiting their relevance in health equity research. What should a disparity measure for the outcome condition on or standardize over? Surveillance reports that track health disparities usually adjust for age and sex, but some decomposition estimators adjust for all covariates needed to identify a causal effect. Ideally, disparity measures should reflect judgements about what constitutes an inequitable difference in the outcome.

Estimators have also ignored how hypothetical interventions are defined. As such, the hypothetical intervention, meant to remove disparities in a targeted factor, may not reflect equity concerns that actual interventionists would respect. For example, an intervention to remove disparities in healthcare should depend on clinical status, but not socioeconomic status which is irrelevant for medical care. Hypothetical interventions should reflect judgements about what constitutes an inequitable difference in the target.

In this paper, we outline a framework for defining disparities and interventions in causal decomposition analysis and provide estimands, non-parametric formulae, and weighting estimators to implement it. Specifically, we draw from the bioethics, biostatistics, epidemiology, and health services research literatures to consider when covariates are 'allowable' for adjusting disparity measures and hypothetical interventions. The estimands, formulae, and estimators we propose partition covariates into 'allowable' sets that define the disparity and intervention, and a 'non-allowable' set that is needed to identify the causal effect. Under a motivating example, we use this framework to examine estimators of natural and interventional effects (mediation analysis) and discrimination (Oaxaca-Blinder decomposition) and the equity judgements they imply and consider a meaningful alternative.

**Motivating Example and Notation**

To fix ideas, we consider a motivating example from clinical medicine: how to reduce disparities in hypertension control by intervening on decisions to intensify antihypertensive treatment.[9] Suppose a healthcare system administrator wants to address disparities in hypertension control across race/ethnicity and tasks us with forming a cohort to study them. Patients are enrolled at their first visit if, on the basis of their systolic blood pressure ($L_1$), they can be classified as hypertensive ($Y_1$;1=yes, $L_1 \geq 140$ mm Hg, 0 =no, $L_1 < 140$ mm Hg). For simplicity we ignore diastolic blood pressure. At six months follow-up systolic blood pressure ($L_2$) and uncontrolled hypertension ($Y_2$;1=yes, $L_2 \geq 140$ mm Hg, 0 =no, $L_2 < 140$ mm Hg) are measured. Disparities in hypertension control may arise through clinical uncertainty.[10] When providers know less about their patients' medical condition (due to poor provider communication, for example) their decision-making may rely on stereotypes. We are interested in how eliminating disparities in treatment decisions would affect disparities in hypertension control.

We thus record the patient's race $R_0$ where $R_0 = r_0$ represents membership in the marginalized group (e.g., blacks) and $R_0 = r_0'$ the privileged group (e.g., whites), demographics age $X_0^{age}$ and sex $X_0^{sex}$, whether antihypertensive treatment was intensified at the initial visit ($M_1$; 1=yes,0=no) as well as socioeconomic factors such as educational attainment $X_0^{edu}$ and private health insurance $X_0^{ins}$ that may implicitly affect treatment and hypertension control. We also record diabetes diagnosis $X_0^{dia}$ which, as a marker of cardiovascular risk, predicts blood pressure and may influence treatment decisions.

The subscripts in our notation grossly indicates a temporal ordering, where '0' indicates variables that are realized before baseline, '1' indicate those realized at baseline, and '2' indicates those realized at follow-up. We assume that some associations between race/ethnicity $R_0$, $X_0$, and $L_1$ are driven by historical processes $H_0$ (e.g., slavery, Jim Crow, federal and local housing policies) as depicted in our causal graph (Figure 1), provided for intuition. We allow that unmeasured factors $U_0$ may correlate repeated measures of systolic blood pressure $L_1$ and $L_2$ and possibly other variables but do not independently predict treatment intensification $M_1$. In our notation, random variables are written in uppercase and their realizations in lowercase. For any variable $V$, the probability $P(V = v|W = w)$ is abbreviated as $P(v|w)$.

To develop general formulae, we will discuss two non-overlapping sets of 'allowable' covariates, those used to define the disparity in hypertension control (the outcome $Y_2$), denoted as $\boldsymbol{A_1^y}$, and those used to further define the intervention on treatment intensification (the targeted factor $M_1$), denoted as $\boldsymbol{A_1^m}$. These sets are restricted to variables not affected by treatment intensification $M_1$. While the time subscript on these sets indicates measurement at baseline, they can include variables measured before baseline as well.

We let $G_{m_1|a_1^y,a_1^m}$ denote a hypothetical stochastic intervention[11,12] to set the conditional distribution of treatment intensification $M_1$ among blacks to the distribution among whites with identical values for $\boldsymbol{A_1^y}$ and $\boldsymbol{A_1^m}$, denoted as $P(m_1|R_0 = r_0', \boldsymbol{a_1^y}, \boldsymbol{a_1^m})$. Suppose we choose $\boldsymbol{A_1^y}$ as $X^{age}$ and $X^{sex}$, $\boldsymbol{A_1^m}$ as $X^{dia}$ and $L_1$. Among blacks, under $G_{m_1|a_1^y a_1^m}$, their treatment is intensified according to a random draw from the distribution among whites who share the same values for $X_0^{age}$, $X_0^{sex}$, $X_0^{dia}$ and , $L_1$.

**Meaningful Disparity and Intervention Definition**

The disparity in hypertension control arises in large part because contextual and other cardiovascular risk factors, and also access to and quality of medical care, are differentially distributed across race.[13] In the U.S., these inequities result from centuries of injustice, including slavery, Jim Crow, government policy, and other forms of structural, cultural, and personal racism.[14] In our example, causal decomposition analysis asks how, among those with uncontrolled hypertension at baseline, the disparity in hypertension control at follow-up $Y_2$ is affected by intervening to eliminate the disparity in treatment decisions $M_1$.

Defining a disparity is a complex process involving decisions about what is fair and just in the distribution of health and its determinants.[15-19] We focus on which covariates are considered 'allowable' for adjustment in defining outcome disparities and interventions, and how these choices relate to equity value judgements. The notion of allowability has been discussed in the context of medical goods and used to define healthcare disparities in 'counterfactual' terms.[20,21] Our discussion broadens the concept to health disparities and decomposition analysis but avoids counterfactual definitions of disparity.

*Allowability*

In the bioethics literature, many define a health disparity as an avoidable, systematic difference between socially advantaged vs. marginalized groups, wherein the marginalized group is further disadvantaged on health.[22,23] In the health services research literature, disparities in healthcare are often defined using the Institute of Medicine definition, as differences in healthcare services that are not due to differences in underlying health needs or preferences.[20,24] A careful reading of these definitions recognizes that, in defining disparities, both avoid detailing a causal model for how they arise. While there are some objections to this,[21] there are practical and scientific reasons why this may be desirable.[22,23]

These definitions encourage us to consider what sources of difference might be considered fair or 'allowable' and to take these 'off the table' when measuring disparity. By corollary, 'non-allowable' sources are those that are unjust and thus contribute to disparity. For health outcomes, allowable sources typically include demographic factors such as age and sex.[25] For medical goods, allowable sources typically include clinical status, history, and presentation.[20,24] All other factors are often considered non-allowable. Although these represent default choices in many studies, other positions become clear when issues of modifiability, amenability to intervention, social contract, and purpose are considered.

*Modifiability & Amenability to Intervention*

In defining health disparity measures, it is common to see innate 'non-modifiable' factors such as chronological age, sex, and even somatic genotype treated as allowable and adjusted for. While society has profound power over the historical distribution of innate factors (through war, genocide, racism, and policy), for a fixed living population, it has no ability to modify them. On these grounds, some might argue that innate differences leading to differences in health are not necessarily unfair.[19]

Some object to using strict modifiability to decide whether to treat a covariate as allowable. The conceptualization and role of innate characteristics in daily, civic, and economic life are entirely under society's control.[26] Moreover, while society cannot change characteristics defined at birth, it can address their effects.[16,22] Social programs can be made age and sex appropriate, increasing their effectiveness. Targeted cancer therapies have been developed for genetic profiles. If one considers a disparity as any difference placing a marginalized group at further disadvantage, and society can address the effects of innate characteristics, one might then consider their differences as contributors to disparity when society fails to adequately respond to them. Thus, innate differences could be considered non-allowable and not adjusted for. The decision to not adjust for innate factors is best applied when the marginalized group is disadvantaged on the innate factor. Failing to adjust otherwise could mask unjust differences.

Consider these arguments in our motivating example. First, we have pre-existing conditions. Blacks disproportionately encounter barriers to care, such as lack of health insurance, leading to higher prevalence of chronic conditions at baseline encounters. By that point, a patient's clinical *history* is beyond the control of the clinician and healthcare system. However, this history can be managed for better prognosis, for example by consulting a specialist. Clinical guidelines recommend tailored treatment protocols for patients with diabetes, kidney disease, and heart failure. Thus, healthcare can respond to disparate pre-existing conditions. Failure to do so equitably would contribute to unjust differences in prognosis. Second, we have age. Blacks are typically younger than whites, and increasing age predicts poor hypertension control. Not adjusting for age would mask unjust differences in hypertension control. Overall, when measuring disparities in prognosis, if there are disparities in pre-existing conditions, one might reasonably decide to treat them as non-allowable (and not adjust). If blacks are younger than whites, one might reasonably decide to treat age as allowable (and adjust).

*Social Contract*

When defining hypothetical interventions that address disparities in goods, allowability choices should consider the social contract. The distribution of any good is ideally governed by norms and conditions that society has agreed upon as fair and just. That is, we believe that decisions about goods are fair if and only if they are based upon ideal criteria that reflect our shared norms and values. That is, in defining interventions on goods, we treat our idealized criteria as allowable.

These arguments have clear implications for interventions that address medical goods. Medical ethics dictates that decisions be clinically appropriate. For addressing disparities in diagnosis, allowable sources could include information needed to accurately differentiate between syndromes, such as presentation and test results. For addressing disparities in treatment, allowable sources could include factors that indicate and modify treatment effectiveness, such as comorbid conditions.[24] For addressing disparities in social conditions, perhaps through a community health worker, allowable sources could include social needs. In our motivating example, the hypothetical intervention to intensify treatment would need to consider age, sex, baseline blood pressure, and diabetes but ignore socioeconomic status. Otherwise, the intervention would not only be unethical, but also inequitable. It would preserve racial differences in treatment that operate through racial differences in socioeconomic status.[5] Observing the social contract reflects that our goal is not equal treatment, but equitable treatment.[24]

When defining disparities in outcomes that are goods, allowability choices should also consider the social contract. This again would treat idealized decision-inputs as allowable. An exception can be made when one seeks to reduce racial differences in goods by eliminating disparities in a criterion governing their distribution. Suppose one were studying a lower rate of listing for transplantation among blacks vs. whites. Listing decisions often consider anticipated social support. An overall racial difference in listing partly due to differences in social support might be concerning, even considered disparate. Here, social support could be treated as non-allowable (and not adjusted for) so it could be studied as a hypothetical intervention.

*Purpose*

In defining outcome disparity measures, allowability must also consider their use. In surveillance and quality assessment, disparity measures can track how well a society, institution, or actor meets objectives. In these settings, when factors the actor does not control are treated as non-allowable (and not adjusted for), this can lead to deleterious effects. In our example, if an external body were benchmarking clinical practices based on disparities in hypertension control rates without any risk adjustment, those that serve marginalized populations with comorbidities may score worse. Those clinics might then be incentivized to avoid complex patients.[27] When causal decomposition analysis is applied to performance assessment measures, it is advisable to study the one used in practice, which might call for pre-existing conditions to be treated as allowable, for the sake of risk adjustment.[28]

*Measurement*

We now turn to how outcome disparities are defined in our causal decomposition analysis. For the outcome, we define disparity as the mean outcome difference across levels of social groups, where the distribution of allowable covariates is standardized. Here, we use the pooled distribution as the standard. In defining the hypothetical intervention, which is intended to remove disparities in a targeted factor, we adjust for allowable covariates through conditioning. We have chosen these simple statistical measures of disparity because the observed disparities can be estimated directly from the data with minimal assumptions and are consistent with widely adopted definitions of disparity.[22-24] These definitions assume common support for the *allowable* covariates across race. Otherwise adjustment would fail to remove the influence of allowable covariates. Defining disparities using the distribution among blacks or whites, rather than the pooled as the standard may weaken this common support assumption.

**Meaningful Disparity Decomposition**

Here, we present general causal decomposition estimands defined by allowable covariates for the disparity in hypertension control (the outcome $Y_2$) and the hypothetical intervention to intensify treatment (the targeted intervention $M_1$). Under assumptions, we provide identifying formulae and weighting-based estimators. Importantly, these formulae and estimators can incorporate confounders that are considered non-allowable, without using them to define outcome disparities or the intervention. All expressions condition on the population of interest, persons with uncontrolled hypertension at baseline.

*Definition*

The observed disparity in uncontrolled hypertension is:

$$\sum_{a_1^y} E[Y_2|r_0, \boldsymbol{a_1^y}] P(\boldsymbol{a_1^y}) - \sum_{a_1^y} E[Y_2|r_0', \boldsymbol{a_1^y}] P(\boldsymbol{a_1^y}) \qquad (1)$$

The change in disparity under the intervention $G_{m_1|a_1^y, a_1^m}$ to remove the disparity in treatment intensification is:

$$\sum_{a_1^y} E[Y_2|r_0, a_1^y]P(a_1^y) - \sum_{a_1^y} E\left[Y_2\left(G_{m_1|a_1^y,a_1^m}\right)\Big|r_0, a_1^y\right]P(a_1^y) \qquad (2)$$

The remaining disparity after the intervention $G_{m_1|a_1^y,a_1^m}$ is:

$$\sum_{a_1^y} E\left[Y_2\left(G_{m_1|a_1^y,a_1^m}\right)\Big|r_0, a_1^y\right]P(a_1^y) - \sum_{a_1^y} E[Y_2|r_0', a_1^y]P(a_1^y) \qquad (3)$$

The covariates $A_1^y$ used to define the disparity are considered outcome-allowable. These covariates $A_1^y$, along with $A_1^m$, are used to define the intervention so both are considered target-allowable. These formulae (1)-(3) are agnostic about whether components of $A_1^y$ affect components of $A_1^m$ and vice-versa, and are also agnostic about whether $A_1^y$ and $A_1^m$ are affected by race $R_0$, but do require that $A_1^y$ and $A_1^m$ are not affected by the targeted variable, treatment intensification $M_1$. This will be satisfied in a design where the eligibility criteria defining the population of interest, the timing of the intervention on $M_1$, and the start of follow-up coincide, and allowable covariates are measured at or before this moment.

*Identification*

The counterfactual disparity reduction (2) and residual (3) are not observed. These expressions can be identified with observational data under assumptions (see eAppendix for formal statements). Among blacks $R_0 = r_0$, we assume conditional exchangeability,[29] or no unmeasured confounding of the relationship between treatment intensification $M_1$ and hypertension control $Y_2$: given the allowables $A_1^y$ and $A_1^m$ and an additional set of non-allowable confounders $N_1$, the potential outcomes under treatment intensification are independent of observed treatment intensification. We further assume positivity[29] among blacks, where there is a positive conditional probability of each observed value for treatment intensification $M_1$ given the allowable covariate sets $A_1^y$ and $A_1^m$ used to define the estimands and the non-allowable covariates $N_1$ used to help identify them. Subtly, it is sufficient if these conditions hold only for the values of $M_1$ observed among whites who share (with blacks) identical values of $A_1^y$ and $A_1^m$. As a consequence, they can be satisfied when, for some subgroups, all have their treatment intensified. We also assume common support across race $R_0$ for the targeted factor and allowable covariates (jointly). Finally, we assume consistency[29] among blacks, that their outcomes would be the same regardless if their values were merely observed or set by hypothetical intervention. These assumptions are strong and the ability to satisfy them will vary across substantive settings.[30]

When these assumptions hold, we can identify the proportion with uncontrolled hypertension among blacks under $G_{m_1|a_1^y,a_1^m}$ as:

$$\sum_{a_1^y} E\left[Y_2\left(G_{m_1|a_1^y,a_1^m}\right)\Big|r_0, a_1^y\right] P(a_1^y)$$

$$= \sum_{m_1,n_1,a_1^m,a_1^y} E[Y_2|r_0, m_1, n_1, a_1^m, a_1^y] P(m_1|r_0', a_1^m, a_1^y)P(n_1|r_0, a_1^m, a_1^y)P(a_1^m|r_0, a_1^y)P(a_1^y) \quad (4)$$

This allows us to identify the disparity reduction (2) given that the observed proportion of uncontrolled hypertension among blacks is:

$$\sum_{a_1^y} E[Y_2|r_0, a_1^y]P(a_1^y)$$

$$= \sum_{m_1,n_1,a_1^m,a_1^y} E[Y_2|r_0, m_1, n_1, a_1^m, a_1^y] P(m_1|r_0, n_1, a_1^m, a_1^y)P(n_1|r_0, a_1^m, a_1^y)P(a_1^m|r_0, a_1^y)P(a_1^y) \quad (5)$$

We can also identify the disparity residual (3) given that the observed proportion of uncontrolled hypertension among whites is:

$$\sum_{a_1^y} E[Y_2|r_0', a_1^y]P(a_1^y)$$

$$= \sum_{m_1,a_1^m,a_1^y} E[Y_2|r_0', m_1, a_1^m, a_1^y] P(m_1|r_0', a_1^m, a_1^y)P(a_1^m|r_0', a_1^y)P(a_1^y) \quad (6)$$

Note that in equations (4)-(6) $P(a_1^y)$ is replaced by $P(a_1^y|r_0)$ if the distribution of outcome-allowable covariates that standardizes the disparity measure is drawn from blacks, and by $P(a_1^y|r_0')$ if drawn from whites.

*Estimation*

Equations (4)-(6) could be estimated using Monte Carlo integration similar to the parametric g-formula,[31] using correctly specified parametric models, as in other contexts.[32-37] A key distinction from these proposals is what the models condition on. The outcome models condition on: (a) allowable and non-allowable covariates in the observed and counterfactual scenarios for blacks (b) only allowable covariates in the observed scenario among whites. The target factor models condition on: (a) allowable and non-allowable covariates in the observed scenario for blacks (b) only allowable covariates in the observed scenario for whites and the counterfactual scenario for blacks. Other algorithms are possible (see eAppendix) but this one does not require non-allowables to be measured or even defined among whites.

Because simulation-based approaches are computationally intensive and require correctly specifying several models, we present two simple weighting-based estimators that can be implemented with standard statistical software routines (briefly described in the e-Appendix). In the health equity context, these encompass existing estimators in the economics,[38-40] sociology,[41,42] biostatistics and epidemiology literatures,[43-48] including those used to estimate natural direct and indirect effects,[49,50] path-specific effects,[51] interventional effects,[11,12,46] and discrimination[52,53] under certain allowability choices, and novel adaptations under others.

Ratio of Mediator Probability Weighting Estimation (RMPW)

The first weighting procedure is based on a ratio of probabilities for treatment intensification $M_1$. The disparity reduction (2) and residual (3) under the intervention $G_{m_1|a_1^y, a_1^m}$ are estimated by comparing weighted means for blacks and whites. First, we estimate the observed proportion with uncontrolled hypertension $Y_2$ among blacks, standardized for the outcome-allowable covariates $A_1^y$:

$$\sum_{a_1^y} E[Y_2|r_0, a_1^y] P(a_1^y) = E[Y_2 \times w_{r_0}|r_0, a_1^y]|r_0] \tag{7a}$$

where the weight $w_{r_0} = \frac{P(r_0)}{P(r_0|a_1^y)}$ \hfill (7b)

Next, we estimate the observed proportion with uncontrolled hypertension $Y_1$ among whites, standardized for the outcome-allowable covariates $\boldsymbol{A_1^y}$:

$$\sum_{\boldsymbol{a_1^y}} E[Y_2|r_0', \boldsymbol{a_1^y}] P(\boldsymbol{a_1^y}) = E[Y_2 \times w_{r_0'}|r_0', \boldsymbol{a_1^y}]|r_0'] \tag{8a}$$

where the weight $w_{r_0'} = \dfrac{P(r_0')}{P(r_0'|\boldsymbol{a_1^y})}$ \hfill (8b)

Last, we estimate the counterfactual proportion with uncontrolled hypertension $Y_2$ among blacks under the intervention $G_{m_1|\boldsymbol{a_1^y},\boldsymbol{a_1^m}}$, which depends on the outcome- and target-allowable covariates $\boldsymbol{A_1^y}$ and $\boldsymbol{A_1^m}$. Like the observed outcomes, the estimated counterfactuals are standardized for the outcome-allowable covariates $\boldsymbol{A_1^y}$:

$$\sum_{\boldsymbol{a_1^y}} E\left[Y_2(G_{m_1|\boldsymbol{a_1^y},\boldsymbol{a_1^m}})\Big|r_0, \boldsymbol{a_1^y}\right] P(\boldsymbol{a_1^y}) = E[E[Y_2 \times w_{r_0}^{rmpw}|r_0, m_1, \boldsymbol{n_1}, \boldsymbol{a_1^m}, \boldsymbol{a_1^y}]|r_0] \tag{9a}$$

where the weight $w_{r_0}^{rmpw} = \dfrac{P(m_1|r_0', \boldsymbol{a_1^m}, \boldsymbol{a_1^y})}{P(m_1|r_0, \boldsymbol{n_1}, \boldsymbol{a_1^m}, \boldsymbol{a_1^y})} \times \dfrac{P(r_0)}{P(r_0|\boldsymbol{a_1^y})}$ \hfill (9b)

The counterfactual disparity reduction is obtained by subtracting (9a) from (7a), and the residual by subtracting (8a) from (9a).

Inverse Odds Ratio Weighting Estimation (IORW)

The second weighting procedure is based on a ratio of inverted odds for race $R$. The disparity reduction (2) and residual (3) under the intervention $G_{m_1|\boldsymbol{a_1^y},\boldsymbol{a_1^m}}$ are estimated by comparing weighted means for blacks and whites. The observed proportions with uncontrolled hypertension $Y_1$ among black and whites, standardizing for the outcome-allowable covariates $\boldsymbol{A_1^y}$, are respectively given in (7a) and (8a). The counterfactual proportion with uncontrolled hypertension $Y_2$ among blacks under the intervention $G_{m_1|\boldsymbol{a_1^y},\boldsymbol{a_1^m}}$, standardized for the outcome-allowable covariates $\boldsymbol{A_1^y}$, is given by:

$$\sum_{\boldsymbol{a_1^y}} E\left[Y_2(G_{m_1|\boldsymbol{a_1^y},\boldsymbol{a_1^m}})\Big|r_0, \boldsymbol{a_1^y}\right] P(\boldsymbol{a_1^y}) = E[E[Y_2 \times w_{r_0}^{iorw}|r_0, m_1, \boldsymbol{n_1}, \boldsymbol{a_1^m}, \boldsymbol{a_1^y}]|r_0] \tag{10a}$$

where the weight $w_{r_0}^{iorw} = \frac{\frac{P(r_0'|m_1, a_1^m, a_1^y)}{P(r_0|m_1, n_1, a_1^m, a_1^y)}}{\frac{P(r_0'|a_1^m, a_1^y)}{P(r_0|n_1, a_1^m, a_1^y)}} \times \frac{P(m_1|a_1^m, a_1^y)}{P(m_1|n_1, a_1^m, a_1^y)} \times \frac{P(r_0)}{P(r_0|a_1^y)}$ (10b)

When all covariates are treated as allowable, the middle term in (10b) cancels and does not need to be estimated. As with RMPW, the disparity reduction is estimated by subtracting (10a) from (7a), and the disparity residual by subtracting (8a) from (10a). Finally, with both RMPW and IORW, each of the weights are multiplied by a factor of $\frac{p(r_0|a_1^y)}{P(r_0)}$ if the distribution of outcome-allowable covariates that standardizes the disparity measure is drawn from blacks, and by a factor of $\frac{p(r_0'|a_1^y)}{P(r_0')}$ if drawn from whites.

**A Closer Look at Existing Estimators**

Estimators of natural and path-specific effects, their analogues, and the Oaxaca-Blinder decomposition are often applied to study disparities. We now use our general expressions to examine these estimators and the equity value judgements they convey (see e-Appendix for a formal discussion). Many of these estimators were developed in the context of mediation analysis to study the effects of exposures and others to measure discrimination.[5,54,55] When used to study disparities, the decomposition estimands they identify are not always meaningful. The problem arises because these of how they adjust for covariates. They implicitly define outcome disparities and interventions in ways that often ignore principles of modifiability, amenability to intervention, social contract, and purpose. Our generalized estimators explicitly consider allowable and non-allowable partitions, so users can be more intentional.

*Natural Indirect Effect Analogue*

Estimators of the natural direct and indirect effects include the regression approaches of Valeri and VanderWeele[56] and Breen et al.,[57] ratio of mediator probability weighting of Hong,[41,42] natural effect models of Lange et al.,[43] inverse odds ratio weighting of Tchetgen Tchetgen,[45] propensity-score weighting of Huber,[58] imputation approaches of Albert[59] and Vansteelandt, Bekaert and Lange,[60] simulation approaches of Imai et al.[33] and Wang et al,[34] and targeted maximum likelihood estimation of Zheng.[48] In our example, these estimate the observed disparity (1) by conditioning the entire analysis on all covariates needed to de-confound the relationship between treatment decisions and hypertension control (or by standardizing across race). This renders all covariates as outcome-allowable. This is problematic because pre-existing diabetes, low educational attainment, and lack of private insurance are risk factors for poor hypertension control and are more common among blacks. Moreover, their effects on hypertension control are amenable to intervention. Thus, treating these pre-existing conditions as outcome-allowable artefactually diminishes the amount of unjust difference to be studied. Furthermore, these approaches estimate the disparity reduction (2) by allowing the hypothetical intervention on treatment intensification to depend on all covariates. This renders all covariates as target-allowable. This is problematic because, by ignoring the social contract, racial differences in treatment intensification that operate through racial differences in educational attainment would persist.

*Path-Specific Effect Analogues*

Estimators of path-specific effects and their interventional analogues have also been applied to study disparities. In our example, all would estimate the observed disparity as (1) by conditioning the entire analysis on age and sex, treating them as outcome-allowable (belonging to $A_1^y$). But they differ in how they would treat covariates affected by race, which in our example are educational attainment, private insurance, diabetes, and baseline blood pressure. The weighting estimator proposed by VanderWeele, Vansteelandt and Robins[46] would treat such covariates as non-allowable (all belonging to $N_1$, leaving $A_1^m$ empty). Therefore, in estimating the disparity reduction as (2), this intervention to set treatment decisions would only depend on age and sex. This is problematic because our society has generally agreed that treatment decisions should depend on clinical needs. Respecting the social contract would assign clinical needs as target-allowable. In contrast, the weighting approach proposed by Zheng and Van Der Laan[48] (and, upon recoding race, that of Miles et al.,[44]) would estimate the disparity reduction (2) by treating all race-affected covariates as allowable (all belonging to $A_1^m$, leaving $N_1$ empty). This intervention to set treatment decisions would depend not only on clinical needs but also educational attainment and private insurance. This is problematic because, as we argued with the natural indirect effect analogue, medical treatment decisions should not hinge on education or private insurance. Finally, a regression-based estimator of Jackson and VanderWeele[5] and a simulation-based estimator proposed by Vansteelandt and Daniel[36] make the same allowability choice as VanderWeele, Vansteelandt and Robins.[46] Additionally, the latter identifies a path-specific effect that does not map to the disparity reduction (2) in the presence target-outcome confounders affected by race, as in our motivating example, but rather the differential impact of two interventions (see eAppendix).

*Oaxaca-Blinder Decompositions*

The Oaxaca-Blinder Decomposition[52,53] has also been applied to study disparities. The approach estimates the disparity reduction (2) by using linear models to regress the outcome on treatment intensification and all covariates and carrying out a 'detailed decomposition' with respect to treatment intensification.[5] This studies a marginal racial difference in hypertension control, with a hypothetical intervention to remove marginal differences in treatment intensification. Effectively, it treats all covariates as neither outcome-allowable nor target-allowable, leaving $A_1^y$ and $A_1^m$ empty. All covariates are treated as non-allowable and included through $N_1$ only to control for confounding. Alternatively, re-weighting estimators[38,39] of the Oaxaca-Blinder decomposition have been widely used. Here, the racial difference in hypertension control conditions on no covariates, leaving $A_1^y$ empty, but the hypothetical intervention on treatment intensification conditions on all covariates, by including them in $A_1^m$. Thus, no covariates are treated as outcome-allowable but all are treated as target-allowable. These allowability assignments share the problems listed for the estimators of natural and path-specific effect analogues.

*A Meaningful Estimator*

In our example, the many existing estimators we examined do not map to meaningful estimands because of the implicit allowability choices they make, which are summarized in Table 1. We could, in adapting the general formulae and weighting expressions, choose to deem age and sex as outcome-and target-allowable (by including them $A_1^y$), respecting principles of modifiability and amenability to intervention, and deem baseline blood pressure and diabetes as exclusively target-allowable (by including them in $A_1^m$), respecting the social contract. Doing so would treat educational attainment and private insurance as non-allowable (by including them in $N_1$), using them to adjust for confounding but not to define the outcome disparity or hypothetical intervention.

**Discussion**

We have shown how causal decomposition analysis can incorporate equity concerns by partitioning covariates into allowable and non-allowable subsets (the latter used for identification). When a covariate is deemed outcome-allowable, its contribution is removed so that we are left with a difference we consider unjust. The hypothetical intervention to remove disparities in a targeted factor is administered within levels of target-allowable covariates, so that the intervention is equitable. We have discussed how allowability choices can consider issues of modifiability, amenability to intervention, social contract, and purpose, reflecting value judgements about equity. We provided generalized non-parametric formulae and weighting-based estimators that are defined in terms of allowable and non-allowable subsets. Last, we discussed when these estimators reduce to existing ones under certain value judgements, unifying and clarifying various approaches from biostatistics, epidemiology, economics, and sociology in the health equity context.

Our proposal has implications for study design in causal decomposition analysis. Researchers should consider variables needed to sensibly measure disparity, and whether these are measured by the start of follow-up with common support among blacks and whites. When our estimators are used, in particular RMPW and formulae (4)-(6) estimated by monte-carlo integration, the non-allowables only need to be defined and measured among blacks. This is important given the way in which some non-allowable constructs, such as racial discrimination, may occur almost exclusively with racial/ethnic minorities.[8]

Regarding disparity definition, by only discussing additive measures of disparity across more vs. less marginalized groups, and by ignoring group size, we implicitly entertained several value judgements.[16,17,22] Our results easily extend to other scales such as the risk ratio or odds ratio. Our approach involved a single axis of disadvantage, but could be extended to study intersectional disparities.[7,8,61] Regarding disparity measurement, selected populations induce correlations between race and outcomes through collider-stratification,[62] as in our example. Because selection occurs pre-target and conditional exchangeability nonetheless holds, we still have causal identification.[4] The impact and interpretation of this important issue for disparity measurement is left for future work.

Regarding estimation, our approach focused on a single target, and continuous and (even non-rare) binary mean outcomes, allowing for race-target, race-covariate, target-covariate, and covariate-covariate interactions. Considering earlier work on RMPW and IORW estimation,[38,39,43,45] and also simulation-based estimation,[32] our approach should extend to multiple targets, distributional outcomes, repeated outcomes, and survival analysis but this is left for future work. Regarding the intervention, we focused on categorical targets. Extensions to continuous targets could involve estimating conditional densities, expanding on earlier work,[32,38,39] but this may prove difficult with several covariates. The weights must be estimated using correctly specified models (see eAppendix). Our focus was on conceptual issues in definition and their relevance for estimation. Future work will consider practical guidance in implementation.

**Conclusion**

We have outlined a framework for incorporating equity concerns into causal decomposition analysis. Our contributions should be of wide interest, particularly when there are baseline differences in non-modifiable factors or when hypothetical interventions concern socially distributed goods.

| Table 1. Allowability Designations of Estimators | $A_1^y$ | $A_1^m$ | $N_1$ |
|---|---|---|---|
| 1. *Oaxaca-Blinder Decomposition via Linear Models*[52,53]: All covariates are considered non-allowable | ⊘ | ⊘ | $X_0^{age}, X_0^{sex}, X_0^{edu}, X_0^{ins}, X_0^{dia}, L_1$ |
| 2. *Oaxaca-Blinder Decomposition Estimator via Reweighting Functions*: No covariates considered outcome-allowable; all covariates are considered target-allowable[38,39] | ⊘ | $X_0^{age}, X_0^{sex}, X_0^{edu}, X_0^{ins}, X_0^{dia}, L_1$ | ⊘ |
| 3. *Natural Direct/Indirect Effect Analogue Estimators*[33,41-43,48,56-58]: All covariates are considered outcome- and target-allowable | $X_0^{age}, X_0^{sex}, X_0^{edu}, X_0^{ins}, X_0^{dia}, L_1$ | ⊘ | ⊘ |
| 4. *Path-Specific Effect Analogue (I & II) Estimator*[5,36,46]: Demographic covariates considered outcome- and target-allowable; remaining covariates considered non-allowable | $X_0^{age}, X_0^{sex}$ | ⊘ | $X_0^{edu}, X_0^{ins}, X_0^{dia}, L_1$ |
| 5. *Path-Specific Effect Analogue (III) Estimator*[44,47]: Demographic covariates outcome- and target-allowable; remaining covariates considered target-allowable | $X_0^{age}, X_0^{sex}$ | $X_0^{edu}, X_0^{ins}, X_0^{dia}, L_1$ | ⊘ |
| 6. *A Meaningful Estimator*: Demographic covariates considered outcome- and target-allowable; clinical covariates considered target-allowable; socioeconomic covariates considered non-allowable | $X_0^{age}, X_0^{sex}$ | $X_0^{dia}, L_1$ | $X_0^{edu}, X_0^{ins}$ |

Abbreviations: $A_1^y$ outcome- (and target-) allowable covariates, $A_1^m$ exclusively target-allowable covariates, $N_1$ non-allowable covariates, ⊘ the empty set; List of covariates include $X^{age}$ age, $X^{sex}$ sex, $X^{edu}$ educational attainment, $X^{ins}$ private health insurance, $X^{dia}$ diabetes, $L_1$ baseline blood pressure

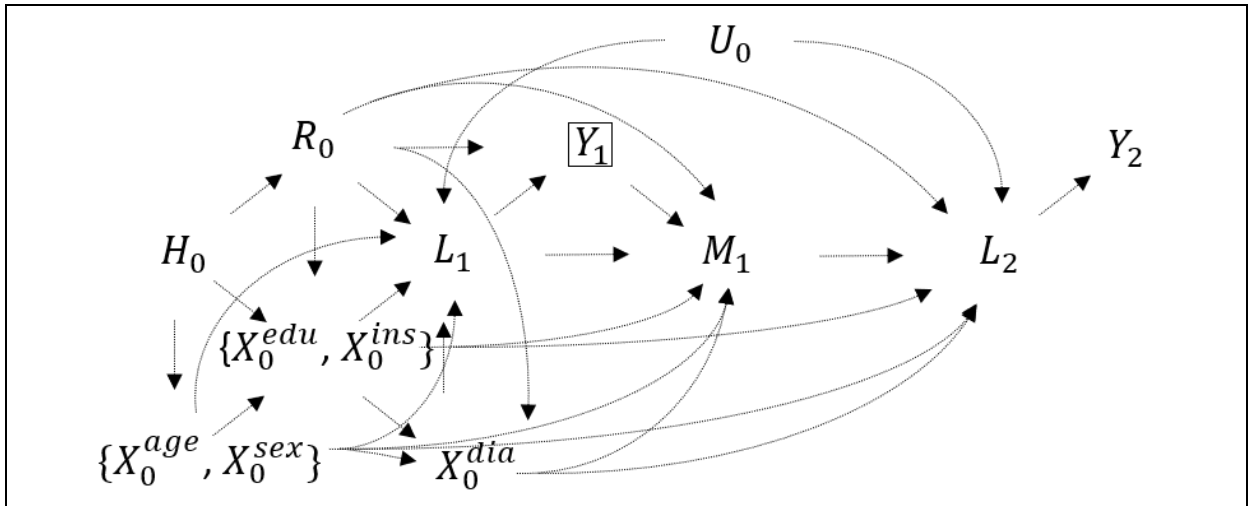

**Figure 1.** Causal diagram depicting relationships between history ($H_0$), race ($R_0$), demographics age and sex ($X^{age}$ and $X^{sex}$), socioeconomic covariates educational attainment and private health insurance ($X^{edu}$ and $X^{ins}$), diabetes ($X^{dia}$), baseline and follow-up blood pressure ($L_1$ and $L_2$), hypertensive status at baseline and subsequent control ($Y_1$ and $Y_2$), and treatment intensification $M_1$. Additional arrows could be allowed from $U$ to the covariates $X$. The subscripts denote, grossly, the time of realization: '0' pre-baseline, '1' baseline, '2' follow-up. The box around $Y_1$ indicates that the population of interest is those with uncontrolled hypertension at baseline. To simplify the graph, brackets group covariates with similar relationships. Within the subset $\{X_0^{age}, X_0^{sex}\}$ no causal relationship is specified; within the subset $\{X_0^{edu}, X_0^{ins}\}$ $X_0^{edu}$ causes $X_0^{ins}$.

**eAppendix.**

Table of Contents

Proofs

**Notation.**

Let the subscript $t$ index the timing of measurement for variable $V_t$ (0=pre-baseline, 1=baseline, 2=follow-up). Let $L_1$ and $L_2$ equal, respectively, a patient's outcome (e.g., blood pressure) at the baseline and follow-up visit. Let $Y_1$ and $Y_2$ equal, respectively, a patient's diagnosis based on $L$ (e.g., uncontrolled hypertension; 1=yes,0=no) at the baseline and follow-up visit. Let $M_1$ equal a determinant of $L_2$ that we want to intervene upon to alter the distribution of $Y_2$ (e.g., decision to intensify antihypertensive treatment; 1=yes,0=no). Let $X_0^{edu}, X_0^{ins}, X_0^{dia}$ (educational attainment, private health insurance, and diabetes, respectively) be measured common causes of $L_1$, $M_1$, and $L_2$, let $X_0^{age}$ and $X_0^{sex}$ (age and sex, respectively) equal common causes of all these variables, let $R_0$ equal a binary variable that defines a socially marginalized population (e.g., race), let $H_0$ equal sociopolitical forces (e.g., racism) that creates association between $R_0$ and $X$. Let $U_0$ equal an unmeasured source of correlation between $L_1$ and $L_2$. (Our results still hold even if this unmeasured cause affects the covariates $X$). For intuition, see eFigure 1 for a causal graph relating these variables. Let $V(w)$ equal the value that $V$ would take (i.e. potential outcome, counterfactual) had $W$ been set to value $w$. Let the notation $V \amalg W | Z$ denote statistical independence between $V$ and $W$ given $Z$.

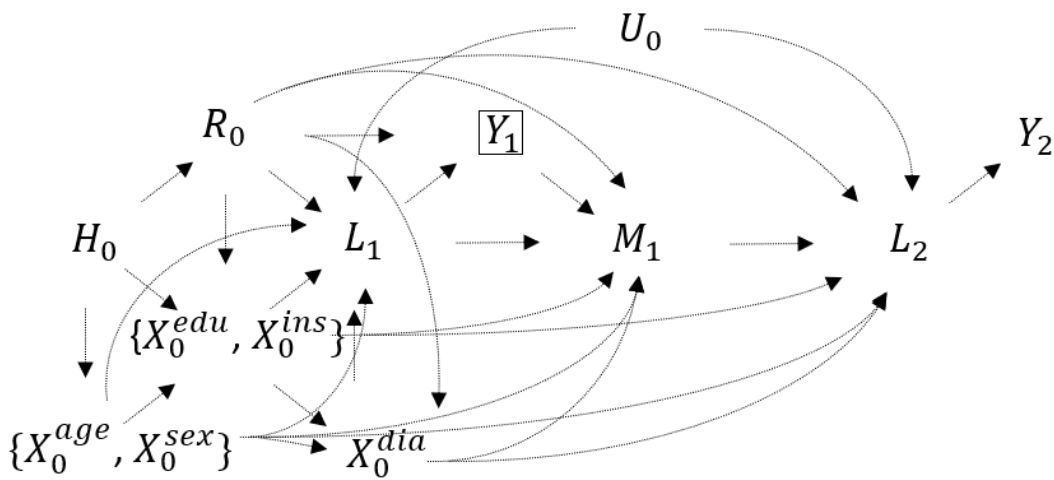

**eFigure 1**. Causal graph describing the the $R_0$—$Y_2$ association through $H, X, L_1, Y_1, M_1,$ and $L_2$

**Definition.**

General formulation.

As defined above, the variable $R$ represents the social status across which the disparity will be measured (e.g. race). $R_0 = r_0$ will represent a marginalized group (e.g. blacks) and $R_0 = r_0'$ the privileged group (e.g. whites). (It is entirely possible to consider the following proposition with these values switched). The population of interest consists of all patients with uncontrolled hypertension at baseline ($Y_1 = 1$). Consider an intervention to set the distribution of a target variable $M_1$ (antihypertensive treatment intensification) to affect disparities in the outcome $Y_2$, uncontrolled hypertension. We will define three non-overlapping sets. The first variable set $\boldsymbol{A_1^y}$ defines the covariates that are considered both outcome- and target-allowable. The second variable set $\boldsymbol{A_1^m}$ defines covariates that are additionally considered target-allowable but not outcome-allowable. The third variable set $\boldsymbol{N_1}$ defines covariates that, in addition to those in $\boldsymbol{A_1^y}$ and $\boldsymbol{A_1^m}$, are needed for causal identification but are nonetheless considered non-allowable. It is permissible to partition the covariates such that some sets remain empty. For example, if $\boldsymbol{A_1^y}$ contains all covariates, then by definition $\boldsymbol{A_1^m}$ and $\boldsymbol{N_1}$ are empty. All expressions that follow condition on the population of interest, patients with hypertension at baseline.

*Proposition.*

Consider an intervention $G_{m_1|a_1^m a_1^y}$ among those with $R_0 = r_0$ to set the distribution of $M_1$ according to the observed distribution $P(m_1|R_0 = r_0', a_1^m, a_1^y)$. The observed disparity prior to intervention, and the reduced and residual disparity after intervention are given, respectively, as:

i) $\sum_{a_1^y} E[Y_2|R_0 = r_0, a_1^y]P(a_1^y) - \sum_{a_1^y} E[Y_2|R_0 = r_0', a_1^y]P(a_1^y)$

ii) $\sum_{a_1^y} E[Y_2|R_0 = r_0, a_1^y]P(a_1^y) - \sum_{a_1^y} E[Y_2(G_{m_1|a_1^m a_1^y})|R_0 = r_0, a_1^y]P(a_1^y)$

iii) $\sum_{a_1^y} E\left[Y_2(G_{m_1|a_1^m a_1^y})\middle|R_0 = r_0, a_1^y\right]P(a_1^y) - \sum_{a_1^y} E[Y_2|R_0 = r_0', a_1^y]P(a_1^y)$

These definitions require that $P(R_0 = r_0|a_1^y) > 0$ and $P(R_0 = r_0'|a_1^y) > 0$ for all $a_1^y$ with $P(a_1^y) > 0$.

*Remark 1.* Alternate definitions of (i)-(iii) that replace the pooled distribution of the outcome-allowable covariates $P(a_1^y)$ can be used. For example, if the distribution the outcome-allowable covariates among blacks were used, we would replace $P(a_1^y)$ with $P(a_1^y|R_0 = r_0)$, under the weaker assumption that $P(R_0 = r_0'|a_1^y) > 0$ for all $a_1^y$ with $P(a_1^y|R_0 = r_0) > 0$. Likewise, if the distribution of the outcome-allowable covariates among whites were used, we would replace $P(a_1^y)$ with $P(a_1^y|R_0 = r_0')$, under the weaker assumption that $P(R_0 = r_0|a_1^y) > 0$ for all $a_1^y$ with $P(a_1^y|R_0 = r_0') > 0$. Alternatively, one could restrict the population of interest to a region where common support holds, perhaps through eligibility criteria. Then, each component of the formulae above would implicitly condition on the eligible population of interest.

Henceforth, we will develop results using the pooled distribution of outcome-allowable covariates $P(a_1^y)$ to measure disparities. The alternatives we outlined will produce diverging estimates when measures of disparity within levels of the outcome-allowable covariates $A_1^y$ vary across levels of these covariates.

**Identification.**

As stated, let $N_1$ denote additional variables needed for conditional exchangeability beyond $A_1^m$ and $A_1^y$.

*Assumptions.*

A) Conditional exchangeability among $R_0 = r_0$:

$Y_2(m) \amalg m_1 | R_0 = r_0, n_1, a_1^m, a_1^y$ for all values $m_1$ with $P(m_1|R_0 = r_0', a_1^m, a_1^y) > 0$

B1) Positivity among $R_0 = r_0$:

$P(m_1|R_0 = r_0, n_1, a_1^m, a_1^y) > 0$ for $n_1, a_1^m, a_1^y$ with $P(n_1, a_1^m, a_1^y|R_0 = r_0) > 0$ for all values $m_1$ with $P(m_1|R_0 = r_0', a_1^m, a_1^y) > 0$

B2) Common support across $R_0$:

$P(a_1^m, a_1^y|R_0 = r_0) > 0$ if $P(a_1^m, a_1^y|R_0 = r_0') > 0$ and $P(m_1|R_0 = r_0, a_1^m, a_1^y) > 0$ if $P(m_1|R_0 = r_0', a_1^m, a_1^y) > 0$ for $a_1^m, a_1^y$ with $P(a_1^m, a_1^y|R_0 = r_0') > 0$

C) Consistency:

$M_{1,i} = m_{1,i} \Rightarrow Y_{2,i} = Y_{2,i}(m_{1,i})$ for all individuals $i$

*Remark 2.* Assumption "A" is a form of "partial" conditional exchangeability: within levels of allowables and non-allowables among blacks, only potential outcomes indexed by certain target values (those values observed among whites with identical values for allowables) are assumed to be independent of the observed target value. Assumption B1 is a form of "partial" positivity: within levels of allowables and non-allowables among blacks, only certain values of the target (those values observed among whites with identical values for allowables) are assumed to be observed with positive probability. These assumptions allow for identification when, for some values of allowables, blacks' treatment is always intensified if whites' treatment is as well. These conditional exchangeability and positivity assumptions are weaker than their standard versions.

Following Jackson & VanderWeele 2018 and Jackson 2018 we have among those with $R_0 = r_0$:

$$\sum_{a_1^y} E\left[Y_2\left(G_{m_1|a_1^m a_1^y} = m_1\right)\middle| R_0 = r_0, a_1^y\right] P(a_1^y)$$
$$= \sum_{m_1, a_1^y, a_1^m} E\left[Y_2(m_1)\middle| R_0 = r_0, G_{m_1|a_1^m a_1^y} = m_1, a_1^m, a_1^y\right] P\left(G_{m_1|a_1^m a_1^y} = m_1\middle| R_0 = r_0, a_1^m, a_1^y\right) P(a_1^m|R_0 = r_0, a_1^y) P(a_1^y)$$
$$= \sum_{m_1, a_1^y, a_1^m} E[Y_2(m_1)|R_0 = r_0, a_1^m, a_1^y] P\left(G_{m_1|a_1^m a_1^y} = m_1\middle| R_0 = r_0, a_1^m, a_1^y\right) P(a_1^m|R_0 = r_0, a_1^y) P(a_1^y)$$
$$= \sum_{m_1, a_1^y, a_1^m} E[Y_2(m_1)|R_0 = r_0, a_1^m, a_1^y] P(M_1 = m_1|R_0 = r_0', a_1^m, a_1^y) P(a_1^m|R_0 = r_0, a_1^y) P(a_1^y)$$
$$= \sum_{m_1, a_1^y, a_1^m, n_1} E[Y_2(m_1)|R_0 = r_0, n_1, a_1^m, a_1^y] P(M_1 = m_1|R_0 = r_0', a_1^m, a_1^y) P(n_1|R_0 = r_0, a_1^m, a_1^y) P(a_1^m|R_0 = r_0, a_1^y) P(a_1^y)$$
$$= \sum_{m_1, a_1^y, a_1^m, n_1} E[Y_2(m_1)|R_0 = r_0, m_1, n_1, a_1^m, a_1^y] P(M_1 = m_1|R_0 = r_0', a_1^m, a_1^y) P(n_1|R_0 = r_0, a_1^m, a_1^y) P(a_1^m|R_0 = r_0, a_1^y) P(a_1^y)$$
$$= \sum_{m_1, a_1^y, a_1^m, n_1} E[Y_2|R_0 = r_0, m_1, n_1, a_1^m, a_1^y] P(M_1 = m_1|R_0 = r_0', a_1^m, a_1^y) P(n_1|R_0 = r_0, a_1^m, a_1^y) P(a_1^m|R_0 = r_0, a_1^y) P(a_1^y) \quad (1)$$

Where the first (and fourth) equality follow by the total law of probability, the second by definition of $G_{m_1|a_1^m a_1^y}$ as random among $R_0 = r_0$ given $a_1^m$ and $a_1^y$, the third by definition of $G_{m_1|a_1^m a_1^y}$ among $R_0 = r_0$ as a random draw from the distribution $P(M_1 = m_1|R_0 = r_0', a_1^m, a_1^y)$ under assumption B2, the fifth by A and B1, and the sixth by C.

Note that among those with $R_0 = r_0$:

$$\sum_{a_1^y} E[Y_2|R_0 = r_0, a_1^y] P(a_1^y)$$
$$= \sum_{m_1, a_1^y, a_1^m, n_1} E[Y_2|R_0 = r_0, m_1, n_1, a_1^m, a_1^y] P(M_1 = m_1|R_0 = r_0, n_1, a_1^m, a_1^y) P(n_1|R_0 = r_0, a_1^m, a_1^y) P(a_1^m|R_0 = r_0, a_1^y) P(a_1^y) \quad (2)$$

And likewise, among those with $R_0 = r_0'$:

$$\sum_{a_1^y} E[Y_2|R_0 = r_0', a_1^y] P(a_1^y)$$
$$= \sum_{m_1, a_1^y, a_1^m} E[Y_2|R_0 = r_0', m_1, a_1^m, a_1^y] P(M_1 = m_1|R_0 = r_0', a_1^m, a_1^y) P(a_1^m|R_0 = r_0', a_1^y) P(a_1^y) \quad (3)$$

Alternatively, among those with $R_0 = r_0'$:

$$\sum_{a_1^y} E[Y_2|R_0 = r_0', a_1^y] P(a_1^y)$$
$$= \sum_{m_1, a_1^y, a_1^m, n_1} E[Y_2|R_0 = r_0', m_1, n_1, a_1^m, a_1^y] P(M_1 = m_1|R_0 = r_0', n_1, a_1^m, a_1^y) P(n_1|R_0 = r_0', a_1^m, a_1^y) P(a_1^m|R_0 = r_0', a_1^y) P(a_1^y) \quad (3^*)$$

*Remark 3.* Equations (1), (2), and (3) represent g-formulae for decomposition with time-fixed interventions.

*Remark 4.* The distribution $P(M_1 = m_1|R_0 = r_0', a_1^m, a_1^y)$ in (1) could be viewed as a marginalization of $P(M_1 = m_1|R_0 = r_0', n_1, a_1^m, a_1^y)$ over the distribution $P(n_1|R_0 = r_0', a_1^m, a_1^y)$ rather than conditional independence between $M_1$ and $N_1$ given $R_0 = r_0'$, $a_1^m$, and $a_1^y$, as can be seen by contrasting the expressions (3) and (3*). Nonetheless, this distribution $P(M_1 = m_1|R_0 = r_0', a_1^m, a_1^y)$ defines the intervention $G_{m_1|a_1^m a_1^y}$ which, when applied to blacks $R_0 = r_0$, produces independence between $M_1$ and $N_1$ given $a_1^m$, and $a_1^y$.

*Remark 5.* The comparison to existing estimators on pages 9 to 13 are derived under the alternate identification formulae (1), (2), and (3*).

### Estimation.

<u>Decomposition using Ratio of Mediator Probability Weights (RMPW)</u>

Let $M_1$ be categorical with $j$ levels $m_{1j}$.

Among those with $R_0 = r_0$ we have that:

$\sum_{a_1^y} E[Y_2(G_{m_1|a_1^m a_1^y})|R_0 = r_0, a_1^y] P(a_1^y)$
$= \sum_{m_1, a_1^y, a_1^m, n_1} E[Y_2|R_0 = r_0, m_{1j}, n_1, a_1^m, a_1^y] P(M_1 = m_{1j}|R_0 = r_0', a_1^m, a_1^y) P(n_1|R_0 = r_0, a_1^m, a_1^y) P(a_1^m|R_0 = r_0, a_1^y) P(a_1^y)$

$= \sum_{m_1, a_1^y, a_1^m, n_1} E[Y_2|R_0 = r_0, m_{1j}, n_1, a_1^m, a_1^y] P(M_1 = m_{1j}|R_0 = r_0, n_1, a_1^m, a_1^y) P(n_1|R_0 = r_0, a_1^m, a_1^y) P(a_1^m|R_0 = r_0, a_1^y) P(a_1^y)$
$\quad \times \dfrac{P(M_1 = m_{1j}|R_0 = r_0', a_1^m, a_1^y)}{P(M_1 = m_{1j}|R_0 = r_0, n_1, a_1^m, a_1^y)}$

$= \sum_{m_1, a_1^y, a_1^m, n_1} E[Y_2|R_0 = r_0, m_{1j}, n_1, a_1^m, a_1^y] P(M_1 = m_{1j}|R_0 = r_0, n_1, a_1^m, a_1^y) P(n_1|R_0 = r_0, a_1^m, a_1^y) P(a_1^m|R_0 = r_0, a_1^y) P(a_1^y|R_0 = r_0)$
$\quad \times \dfrac{P(M_1 = m_{1j}|R_0 = r_0', a_1^m, a_1^y)}{P(M_1 = m_{1j}|R_0 = r_0, n_1, a_1^m, a_1^y)} \times \dfrac{P(a_1^y)}{P(a_1^y|R_0 = r_0)}$

$= \sum_{m_1, a_1^y, a_1^m, n_1} E[Y_2|R_0 = r_0, m_{1j}, n_1, a_1^m, a_1^y] P(M_1 = m_{1j}|R_0 = r_0, n_1, a_1^m, a_1^y) P(n_1|R_0 = r_0, a_1^m, a_1^y) P(a_1^m|R_0 = r_0, a_1^y) P(a_1^y|R_0 = r_0)$
$\quad \times \dfrac{P(M_1 = m_{1j}|R_0 = r_0', a_1^m, a_1^y)}{P(M_1 = m_{1j}|R_0 = r_0, n_1, a_1^m, a_1^y)} \times \dfrac{P(r_0)}{P(r_0|a_1^y)}$

$= E[E[Y_2 \times w_{r_0}^{rmpw}|r_0, m_1, n_1, a_1^m, a_1^y]|r_0]$  (4a)

where $w_{r_0}^{rmpw} = \dfrac{P(M_1 = m_{1j}|R_0 = r_0', a_1^m, a_1^y)}{P(M_1 = m_{1j}|R_0 = r_0, n_1, a_1^m, a_1^y)} \times \dfrac{P(r_0)}{P(r_0|a_1^y)}$  (4b)

The first equality is identified via eqn 1.

Note that, among those with $R_0 = r_0$ we have:

$\sum_{a_1^y} E[Y_2|r_0, a_1^y] P(a_1^y)$
$= E[E[Y_2 \times w_{r_0}|r_0, m_1, n_1, a_1^m, a_1^y]|r_0]$  (5a)

where $w_{r_0} = \dfrac{P(r_0)}{P(r_0|a_1^y)}$  (5b)

And among those with $R_0 = r_0'$:

$\sum_{a_1^y} E[Y_2|r_0', a_1^y] P(a_1^y)$
$= E[E[Y_2 \times w_{r_0'}|r_0', m_1, a_1^m, a_1^y]|r_0']$  (6a)

where $w_{r_0'} = \dfrac{P(r_0')}{P(r_0'|a_1^y)}$  (6b)

Thus, under the expressions and weights defined above, we have the general result:

The observed disparity
$$\psi^{obs} = \sum_{a_1^y} E[Y_2|R_0 = r_0, a_1^y]P(a_1^y) - \sum_{a_1^y} E[Y_2|R_0 = r_0', a_1^y]P(a_1^y) \tag{7a}$$
$$= E[E[Y_2 \times w_{r_0}|r_0, m_1, \boldsymbol{n_1}, a_1^m, a_1^y]|r_0] - E[E[Y_2 \times w_{r_0'}|r_0', m_1, a_1^m, a_1^y]|r_0']$$

The reduced disparity
$$\psi^{red} = \sum_{a_1^y} E[Y_2|R_0 = r_0, a_1^y]P(a_1^y) - \sum_{a_1^y} E[Y_2(G_{m_1|a_1^m a_1^y})|R_0 = r_0, a_1^y]P(a_1^y) \tag{7b}$$
$$= E[E[Y_2 \times w_{r_0}|r_0, m_1, \boldsymbol{n_1}, a_1^m, a_1^y]|r_0] - E[E[Y_2 \times w_{r_0}^{rmpw}|r_0, m_1, \boldsymbol{n_1}, a_1^m, a_1^y]|r_0]$$

The residual disparity
$$\psi^{res} = \sum_{a_1^y} E[Y_2(G_{m_1|a_1^m a_1^y})|R_0 = r_0, a_1^y]P(a_1^y) - \sum_{a_1^y} E[Y_2|R_0 = r_0', a_1^y]P(a_1^y) \tag{7c}$$
$$= E[E[Y_2 \times w_{r_0}^{rmpw}|r_0, m_1, \boldsymbol{n_1}, a_1^m, a_1^y]|r_0] - E[E[Y_2 \times w_{r_0'}|r_0', m_1, a_1^m, a_1^y]|r_0']$$

With weights defined as
$$w_{r_0} = \frac{P(r_0)}{P(r_0|a_1^y)}$$

$$w_{r_0'} = \frac{P(r_0')}{P(r_0'|a_1^y)}$$

$$w_{r_0}^{rmpw} = \frac{P(M_1 = m_{1j}|R_0 = r_0', a_1^m, a_1^y)}{P(M_1 = m_{1j}|R_0 = r_0, \boldsymbol{n_1}, a_1^m, a_1^y)} \times \frac{P(r_0)}{P(r_0|a_1^y)}$$

Contrast (7a) can be estimated as $\beta_1$ in the weighted regression model with the observed data:

$E[Y_2|R_0] = \beta_0 + \beta_1 R_0$ fit with weights $w_{r_0}$ for those with $R_0 = r_0$ and $w_{r_0'}$ for those with $R_0 = r_0'$.

Constrast (7b) can be estimated as $\beta_1$ in the weighted regression model with a stacked dataset consisting of the original subset $R_0 = r_0$ (labelled as $D_0 = d_0$) and a copy of the subset $R_0 = r_0$ (labelled as $(D_0 = d_0')$).

$E[Y_2|D_0] = \beta_0 + \beta_1 D_0$ fit with weights $w_{r_0}$ for those with $D_0 = d_0$ and $w_{r_0}^{rmpw}$ for those with $D_0 = d_0'$.

Contrast (7c) can be estimated as $\beta_1$ in the weighted regression model with the observed data:

$E[Y_2|R_0] = \beta_0 + \beta_1 R_0$ fit with weights $w_{r_0}^{rmpw}$ for those with $R_0 = r_0$ and $w_{r_0'}$ for those with $R_0 = r_0'$.

*Remark 6.* (7a), (7b), and (7c) are based on disparity measures that use the pooled distribution $P(a_1^y)$ to standardize the outcome-allowable covariates. With $P(a_1^y|R_0 = r_0)$ as the standard the weights would be:

$$w_{r_0} = 1 \qquad w_{r_0'} = \frac{P(r_0|a_1^y)}{P(r_0'|a_1^y)} \times \frac{P(r_0')}{P(r_0)} \qquad w_{r_0}^{rmpw} = \frac{P(M_1 = m_{1j}|R_0 = r_0', a_1^m, a_1^y)}{P(M_1 = m_{1j}|R_0 = r_0, \boldsymbol{n_1}, a_1^m, a_1^y)}$$

With $P(a_1^y|R_0 = r_0')$ as the standard the weights would be:
$$w_{r_0} = \frac{(r_0'|a_1^y)}{P(r_0|a_1^y)} \times \frac{P(r_0)}{P(r_0')} \qquad w_{r_0'} = 1 \qquad w_{r_0}^{rmpw} = \frac{P(M_1 = m_{1j}|R_0 = r_0', a_1^m, a_1^y)}{P(M_1 = m_{1j}|R_0 = r_0, \boldsymbol{n_1}, a_1^m, a_1^y)} \times \frac{(r_0'|a_1^y)}{P(r_0|a_1^y)} \times \frac{P(r_0)}{P(r_0')}$$

*Remark 7.* The conditionality of the intervention $G_{m_1|a_1^m a_1^y}$ appears through the numerator in (4b). Any non-allowable confounders $N_1$ beyond the allowable variables defined in $A_1^m$ and $A_1^y$ appear only in the denominator. Thus, the conditionality of the numerator will differ from the denominator whenever the intervention $G_{m_1|a_1^m a_1^y}$ does not condition on all of the confounders of $M_1$.

Decomposition using Inverse Odds Ratio Weights (IORW)

Let $M_1$ be categorical with $j$ levels $m_{1j}$.

Among those with $R_0 = r_0$ we have that:

$\sum_{a_1^y} E[Y_2(G_{m_1|a_1^m a_1^y})|R_0 = r_0, a_1^y] P(a_1^y)$

$= \sum_{m_1, a_1^y, a_1^m, n_1} E[Y_2|R_0 = r_0, m_{1j}, n_1, a_1^m, a_1^y] P(M_1 = m_{1j}|R_0 = r_0', a_1^m, a_1^y) P(n_1|R_0 = r_0, a_1^m, a_1^y) P(a_1^m|R_0 = r_0, a_1^y) P(a_1^y)$

$= \sum_{m_1, a_1^y, a_1^m, n_1} E[Y_2|R_0 = r_0, m_{1j}, n_1, a_1^m, a_1^y] P(M_1 = m_{1j}|R_0 = r_0, n_1, a_1^m, a_1^y) P(n_1|R_0 = r_0, a_1^m, a_1^y) P(a_1^m|R_0 = r_0, a_1^y) P(a_1^y|R_0 = r_0)$

$\quad \times \dfrac{P(M_1 = m_{1j}|R_0 = r_0', a_1^m, a_1^y)}{P(M_1 = m_{1j}|R_0 = r_0, n_1, a_1^m, a_1^y)} \times \dfrac{P(r_0)}{P(r_0|a_1^y)}$

$= \sum_{m_1, a_1^y, a_1^m, n_1} E[Y_2|R_0 = r_0, m_{1j}, n_1, a_1^m, a_1^y] P(M_1 = m_{1j}|R_0 = r_0, n_1, a_1^m, a_1^y) P(n_1|R_0 = r_0, a_1^m, a_1^y) P(a_1^m|R_0 = r_0, a_1^y) P(a_1^y|R_0 = r_0)$

$\quad \times \dfrac{\dfrac{P(R_0 = r_0', M_1 = m_{1j}, a_1^m, a_1^y)}{P(R_0 = r_0', a_1^m, a_1^y)}}{\dfrac{P(R_0 = r_0, M_1 = m_{1j}, n_1, a_1^m, a_1^y)}{P(R_0 = r_0, n_1, a_1^m, a_1^y)}} \times \dfrac{P(r_0)}{P(r_0|a_1^y)}$

$= \sum_{m_1, a_1^y, a_1^m, n_1} E[Y_2|R_0 = r_0, m_{1j}, n_1, a_1^m, a_1^y] P(M_1 = m_{1j}|R_0 = r_0, n_1, a_1^m, a_1^y) P(n_1|R_0 = r_0, a_1^m, a_1^y) P(a_1^m|R_0 = r_0, a_1^y) P(a_1^y|R_0 = r_0)$

$\quad \times \dfrac{\dfrac{P(R = r_0'|m_{1j}, a_1^m, a_1^y)}{P(R = r_0|m_{1j}, n_1, a_1^m, a_1^y)}}{\dfrac{P(R = r_0'|a_1^m, a_1^y)}{P(R = r_0|n_1, a_1^m, a_1^y)}} \times \dfrac{P(M_1 = m_{1j}|a_1^m, a_1^y)}{P(M_1 = m_{1j}|n_1, a_1^m, a_1^y)} \times \dfrac{P(r_0)}{P(r_0|a_1^y)}$

$= E\big[E[Y_2 \times w_{r_0}^{iorw}|r_0, m_1, n_1, a_1^m, a_1^y]|r_0\big]$ (8a)

where $w_{r_0}^{iorw} = \dfrac{\dfrac{P(R = r_0'|m_{1j}, a_1^m, a_1^y)}{P(R = r_0|m_{1j}, n_1, a_1^m, a_1^y)}}{\dfrac{P(R = r_0'|a_1^m, a_1^y)}{P(R = r_0|n_1, a_1^m, a_1^y)}} \times \dfrac{P(M_1 = m_{1j}|a_1^m, a_1^y)}{P(M_1 = m_{1j}|n_1, a_1^m, a_1^y)} \times \dfrac{P(r_0)}{P(r_0|a_1^y)}$ (8b)

The first equality is identified via eqn 1.

The third and sixth equalities show that $w_{r_0}^{iorw} = w_{r_0}^{rmpw}$ non-parametrically. Thus, we can implement IORW approach by following the procedure outlined with RMPW, replacing $w_{r_0}^{rmpw}$ (4b) by $w_{r_0}^{iorw}$ (8b).

*Remark 8.* The conditionality of the intervention $G_{m_1|a_1^m a_1^y}$ appears through the numerators in (8b). Any non-allowable confounders $N_1$ beyond the allowable variables defined in $A_1^m$ and $A_1^y$ appear only in the denominators. Thus, the conditionality of the numerators will differ from the denominators whenever the intervention $G_{m_1|a_1^m a_1^y}$ does not condition on all of the confounders of $M_1$.

*Remark 9.* The weight in (8b) is based on disparity measures that use the pooled distribution $P(a_1^y)$ to standardize the outcome-allowable covariates. With $P(a_1^y|R_0 = r_0)$ as the standard the weight would be:

$$w_{r_0}^{iorw} = \dfrac{\dfrac{P(R = r_0'|m_{1j}, a_1^m, a_1^y)}{P(R = r_0|m_{1j}, n_1, a_1^m, a_1^y)}}{\dfrac{P(R = r_0'|a_1^m, a_1^y)}{P(R = r_0|n_1, a_1^m, a_1^y)}} \times \dfrac{P(M_1 = m_{1j}|a_1^m, a_1^y)}{P(M_1 = m_{1j}|n_1, a_1^m, a_1^y)}$$

With $P(\boldsymbol{a_1^y}|R_0 = r_0')$ as the standard the weight would be:

$$\frac{\frac{P(R = r_0'|m_{1j}, \boldsymbol{a_1^m}, \boldsymbol{a_1^y})}{P(R = r_0|m_{1j}, \boldsymbol{n_1}, \boldsymbol{a_1^m}, \boldsymbol{a_1^y})}}{\frac{P(R = r_0'|\boldsymbol{a_1^m}, \boldsymbol{a_1^y})}{P(R = r_0|\boldsymbol{n_1}, \boldsymbol{a_1^m}, \boldsymbol{a_1^y})}} \times \frac{P(M_1 = m_{1j}|\boldsymbol{a_1^m}, \boldsymbol{a_1^y})}{P(M_1 = m_{1j}|\boldsymbol{n_1}, \boldsymbol{a_1^m}, \boldsymbol{a_1^y})} \times \frac{P(r_0'|\boldsymbol{a_1^y})}{P(r_0|\boldsymbol{a_1^y})} \times \frac{P(r_0)}{P(r_0')}$$

### Implementation

The sketch for parametric g-computation in the main-text was based on models using the factorizations (1), (2), and (3). However, one can replace (3) with (3*). If that is done, the outcome models for blacks and whites would always condition on allowable and non-allowable covariates. Also, the target factor models would always condition on the allowable and non-allowable covariates among the observed scenarios for blacks and whites, but only condition on the allowables in the counterfactual scenario for blacks. This alternate specification can lead to issues with non-compatibility, as it may be difficult to specify models for $P(M_1 = m_{1j}|R_0 = r_0', \boldsymbol{a_1^m}, \boldsymbol{a_1^y})$ (used for estimating the counterfactual scenario for blacks under (3*)) and $P(M_1 = m_{1j}|R_0 = r_0', \boldsymbol{n_1}, \boldsymbol{a_1^m}, \boldsymbol{a_1^y})$ (used for estimating the observed scenario for whites under (3*)) that are compatible with one another. This challenge does not arise when (3) is used because then only $P(M_1 = m_{1j}|R_0 = r_0', \boldsymbol{a_1^m}, \boldsymbol{a_1^y})$ must be specified.

<u>Ratio of Mediator Probability Weighting</u>

The first component of the weight $w_{r_0}^{rmpw}$ is a ratio of two probabilities. The numerator could be estimated by fitting, among whites, a logistic regression model for the probability of treatment intensification $M_1$ given the allowable covariates $\boldsymbol{A_1^m}$ and $\boldsymbol{A_1^y}$. The denominator could be estimated by fitting an analogous model among blacks that further conditions on non-allowable confounders $\boldsymbol{N_1}$. These models need not be compatible when fitted this way, separately for blacks and whites. The second component of the weight $w_{r_0}^{rmpw}$ could be obtained with logistic regression models for race $R_0$ that do and do not control for the outcome-allowable covariates $\boldsymbol{A_1^y}$. The predicted values from these four models are used to obtain the weight $w_{r_0}^{rmpw}$ for each individual.

A stacking procedure can be used to estimate the effects of interest. To obtain the disparity reduction (2) minus (1), the data from blacks with weight $w_{r_0}$ (5b) are stacked onto a copy from blacks with weight $w_{r_0}^{rmpw}$ (4b) and labelled with a new variable called data origin ($D_0$; 1=original, 0=copy). The weighted mean difference in $Y_2$ across data origin $D_0$ estimates the disparity reduction. To obtain the disparity residual (1) minus (3), the data from blacks with weight $w_{r_0}^{rmpw}$ (4b) are stacked onto the data from whites with weight $w_{r_0'}$ (6b). The weighted mean difference in $Y_2$ across race $R_0$ estimates the disparity residual. For inference, the non-parametric bootstrap could be used to obtain 95% confidence intervals.

<u>Inverse Odds Ratio Weighting</u>

The first component of $w_{r_0}^{iorw}$ is a ratio of two odds. The numerator odds can be estimated by fitting logistic regressions for race $R_0$ given treatment intensification $M_1$, allowable covariates $\boldsymbol{A_1^m}$ and $\boldsymbol{A_1^y}$ with and without further control for non-allowable confounders $\boldsymbol{N_1}$. For the denominator odds one can use similar models but without control for treatment intensification $M_1$. For the second and third components one can adapt what was described for the RMPW-style estimator, with the caveat that the models for treatment intensification $M_1$ do not condition on race $R_0$. As noted in the main text, the estimation procedure is valid if all models are specified correctly, and here special care should be taken to ensure that models are compatible with one another. For guidance, see the procedure proposed by Miles et al. Once all necessary models are fit, their predicted values are used to form individual weights. The stacking procedure described above is used but replacing $w_{r_0}^{rmpw}$ weights (4b) with $w_{r_0}^{iorw}$ weights (8b).

### Relation to Existing Estimators (under identifying formulae (1), (2), and (3*))

*In what follows we make the notation more compact as follows: $X_0^{age} = X_0^g$, $X_0^{sex} = X_0^s$, $X_0^{edu} = X_0^e$, $X_0^{ins} = X_0^i$, $X_0^{dia} = X_0^d$, with sets notated as, e.g., $X_0^g, X_0^s = X_0^{g,s}$. In weight expressions, $M_1$ is categorical with j levels $m_{1j}$.*

Interventional Analogue of the Natural Indirect Effect

Suppose we estimate the disparity reduction where $A_1^y$, the covariates deemed both outcome- and target-allowable, includes all covariates. This leaves $A_1^m$ empty because we have exhausted the potential covariates that could be deemed target-allowable. This leaves $N_1$ empty because we have exhausted the covariates needed to establish conditional exchangeability for $M_1$. The disparity reduction is identified by the non-parametric expression of Pearl and the weighting estimators of Hong (2010) and (2015), Huber, Lange et al., and Tchetgen Tchetgen.

*Non-parametric*

$$\psi^{red} = \sum_{m_1,l_1,x_0^{g,s,e,i,d}} E[Y_2|R_0 = r_0, m_1, l_1, x_0^{g,s,e,i,d}]$$
$$\times \{P(M_1 = m_1|R_0 = r_0, l_1, x_0^{g,s,e,i,d}) - P(M_1 = m_1|R_0 = r_0', l_1, x_0^{g,s,e,i,d})\}$$
$$\times P(l_1, x_0^{g,s,e,i,d})$$

A conditional expression is obtained by removing the integration over $l_1, x_0^{g,s,e,i,d}$. This expression is equivalent to the mediation formula of Pearl, which underlies the regression-based estimators of Valeri and VanderWeele, as well as the simulation-based estimators of Imai et al. and Wang et al. The marginal expression serves as the basis for the imputation estimator of Albert, and VanderWeele and Vansteelandt.

*Ratio of Mediator Probability Weighting*

$$\psi^{red} = E\big[E[Y_2 \times w_{r_0}|r_0, m_1, l_1, x_0^{g,s,e,i,d}]|r_0\big] - E\big[E[Y_2 \times w_{r_0}^{rmpw}|r_0, m_1, l_1, x_0^{g,s,e,i,d}]|r_0\big]$$

Where

$$w_{r_0} = \frac{P(r_0)}{P(r_0|l_1, x_0^{g,s,e,i,d})} \qquad w_{r_0}^{rmpw} = \frac{P(M_1 = m_{1j}|R_0 = r_0', l_1, x_0^{g,s,e,i,d})}{P(M_1 = m_{1j}|R_0 = r_0, l_1, x_0^{g,s,e,i,d})} \times w_{r_0}$$

A conditional expression is obtained by removing the outer expectation and setting $w_{r_0} = w_{r_0'} = 1$. The marginal and conditional versions are equivalent to the weighting approaches of Hong (2010) and (2015) and those used in the natural effect models of Lange et al.

*Inverse Odds Ratio Weighting*

$$\psi^{red} = E\big[E[Y_2 \times w_{r_0}|r_0, m_1, l_1, x_0^{g,s,e,i,d}]|r_0\big] - E\big[E[Y_2 \times w_{r_0}^{iorw}|r_0, m_1, l_1, x_0^{g,s,e,i,d}]|r_0\big]$$

Where

$$w_{r_0} = \frac{P(r_0)}{P(r_0|l_1, x_0^{g,s,e,i,d})} \qquad w_{r_0}^{iorw} = \frac{\frac{P(R = r_0'|m_{1j}, l_1, x_0^{g,s,e,i,d})}{P(R = r_0|m_{1j}, l_1, x_0^{g,s,e,i,d})}}{\frac{P(R = r_0'|l_1, x_0^{g,s,e,i,d})}{P(R = r_0|l_1, x_0^{g,s,e,i,d})}} \times w_{r_0}$$

This is equivalent to the approach of Huber. A conditional expression is obtained by removing the outer expectation and setting $w_r = w_r' = 1$ which is related to a variant of the approach of Tchetgen Tchetgen.

Interventional Analogue of the Path-Specific Indirect Effect I

Suppose we estimate the disparity reduction where $A_1^y$, the covariates deemed both outcome- and target-allowable, include $X^{g,s}$. $A_1^m$ is left empty so that no additional variables are considered target-allowable, and $N_1$ includes all other variables needed to establish conditional exchangeability for $M_1$ (i.e., $L_1, X_0^{e,i,d}$). The disparity reduction is identified by the non-parametric expression and weighting estimator for the interventional indirect effect of VanderWeele, Vansteelandt, and Robins.

*Non-parametric*

$$\psi^{red} = \sum_{m_1, l_1, x_0^{g,s,e,i,d}} E[Y_2|R_0 = r_0, m_1, l_0, x_0^{g,s,e,i,d}]$$
$$\times \{P(M_1 = m_1|R_0 = r_0, l_1, x_0^{g,s,e,i,d}) - P(M_1 = m_1|R_0 = r_0', x_0^{g,s})\}$$
$$\times P(l_1, x_0^{e,i,d}|R_0 = r_0, x_0^{g,s})$$
$$\times P(x_0^{g,s})$$

A conditional expression is obtained by removing the integration over $x_0^{g,s}$. This is equivalent to the expression of VanderWeele, Vansteelandt, and Robins under a stochastic intervention.

*Ratio of Mediator Probability Weighting*

$$\psi^{red} = E\big[E[Y_2 \times w_{r_0}|r_0, m_1, l_1, x_0^{g,s,e,i,d}]|r_0\big] - E\big[E[Y_2 \times w_{r_0}^{rmpw}|r_0, m_1, l_1, x_0^{g,s,e,i,d}]|r_0\big]$$

Where

$$w_{r_0} = \frac{P(r_0)}{P(r_0|x_0^{g,s})} \qquad w_{r_0}^{rmpw} = \frac{P(M_1 = m_{1j}|R_0 = r_0', x_0^{g,s})}{P(M_1 = m_{1j}|R_0 = r_0, l_1, x_0^{g,s,e,i,d})} \times w_{r_0}$$

A conditional expression is obtained by conditioning the outer expectation on $x_0^{g,s}$ and setting $w_r = w_r' = 1$. This is equivalent to the approach of VanderWeele, Vansteelandt, and Robins under a stochastic intervention. They express $P(M_1 = m_{1j}|R_0 = r_0', x_0^{g,s})$ as $\sum_{l_1, x_0^{e,i,d}} P(M_1 = m_{1j}|R_0 = r_0', l_1, x_0^{g,s,e,i,d}) P(l_1, x_0^{e,i,d}|R_0 = r_0', x_0^{g,s})$ to emphasize that $P(M_1 = m_{1j}|R_0 = r_0', x_0^{g,s})$ represents a marginalization of $P(M_1 = m_{1j}|R_0 = r_0', l_1, x_0^{g,s,e,i,d})$ over $P(l_1, x_0^{e,i,d}|R_0 = r_0', x_0^{g,s})$ rather than conditional independence of $M_1$ and $\{l_1, x_0^{e,i,d}\}$ given $R = r_0'$ and $x_0^{g,s}$.

Interventional Analogue of the Path-Specific Indirect Effect II

The simulation-based estimator of the interventional indirect effect of Vansteelandt and Daniel does not generally estimate the disparity reduction, but rather a contrast of two interventions.

To see this, consider two stochastic interventions. The first intervention, $G_{m_1|a_1^m a_1^y}^{r_0}$, assigns treatment intensification, the targeted factor, according to its conditional distribution among blacks given the allowables $A_1^y$ and $A_1^m$, defined as $P(M_1 = m_1|R_0 = r_0, a_1^m, a_1^y)$. The second intervention, $G_{m_1|a_1^m a_1^y}^{r_0'}$, assigns treatment intensification, the targeted factor, according to its conditional distribution among whites given the allowables $A_1^y$ and $A_1^m$, defined as $P(M_1 = m_1|R_0 = r_0', a_1^m, a_1^y)$.

According to assumptions A, a slight variant of B1, and C, under the first intervention, the proportion of blacks with uncontrolled hypertension, standardized by the outcome-allowable covariates, is:

$\sum_{a_1^y} E\left[Y_2\left(G_{m_1|a_1^m a_1^y}^{r_0} = m_1\right) \middle| R_0 = r_0, a_1^y\right] P(a_1^y)$

$= \sum_{m_1, a_1^y, a_1^m} E\left[Y_2(m_1) \middle| R_0 = r_0, G_{m_1|a_1^m a_1^y}^{r_0} = m_1, a_1^m, a_1^y\right] P\left(G_{m_1|a_1^m a_1^y}^{r_0} = m_1 \middle| R_0 = r_0, a_1^m, a_1^y\right) P(a_1^m | R_0 = r_0, a_1^y) P(a_1^y)$

$= \sum_{m_1, a_1^y, a_1^m} E[Y_2(m_1) | R_0 = r_0, a_1^m, a_1^y] P\left(G_{m_1|a_1^m a_1^y}^{r_0} = m_1 \middle| R_0 = r_0, a_1^m, a_1^y\right) P(a_1^m | R_0 = r_0, a_1^y) P(a_1^y)$

$= \sum_{m_1, a_1^y, a_1^m} E[Y_2(m_1) | R_0 = r_0, a_1^m, a_1^y] P(M_1 = m_1 | R_0 = r_0, a_0^m, a_0^y) P(a_1^m | R_0 = r_0, a_1^y) P(a_1^y)$

$= \sum_{m_1, a_1^y, a_1^m, n_1} E[Y_2(m_1) | R_0 = r_0, n_1, a_1^m, a_1^y] P(M_1 = m_1 | R_0 = r_0, a_1^m, a_1^y) P(n_1 | R_0 = r_0, a_1^m, a_1^y) P(a_1^m | R_0 = r_0, a_1^y) P(a_1^y)$

$= \sum_{m_1, a_1^y, a_1^m, n_1} E[Y_2(m_1) | R_0 = r_0, m_1, n_1, a_1^m, a_1^y] P(M_1 = m_1 | R_0 = r_0, a_1^m, a_1^y) P(n_1 | R_0 = r_0, a_1^m, a_1^y) P(a_1^m | R_0 = r_0, a_1^y) P(a_1^y)$

$= \sum_{m_1, a_1^y, a_1^m, n_1} E[Y_2 | R_0 = r_0, m_1, n_1, a_1^m, a_1^y] P(M_1 = m_1 | R_0 = r_0, a_1^m, a_1^y) P(n_1 | R_0 = r_0, a_1^m, a_1^y) P(a_1^m | R_0 = r_0, a_1^y) P(a_1^y)$ (9)

According to equation (1), under assumptions A, B1, B2, and C, under the second intervention, the proportion of blacks with uncontrolled hypertension, standardized by the outcome-allowable covariates, is:

$\sum_{a_1^y} E\left[Y_2\left(G_{m_1|a_1^m a_1^y}^{r_0'} = m_1\right) \middle| R_0 = r_0, a_1^y\right] P(a_1^y)$

$= \sum_{m_1, a_1^y, a_1^m, n_1} E[Y_2 | R_0 = r_0, m_1, n_1, a_1^m, a_1^y] P(M_1 = m_1 | R_0 = r_0', a_1^m, a_1^y) P(n_1 | R_0 = r_0, a_1^m, a_1^y) P(a_1^m | R_0 = r_0, a_1^y) P(a_1^y)$

The difference in uncontrolled hypertension among blacks comparing the two interventions is:

$\sum_{a_1^y} E\left[Y_2\left(G_{m_1|a_1^m a_1^y}^{r_0} = m_1\right) \middle| R_0 = r_0, a_1^y\right] P(a_1^y) - \sum_{a_1^y} E\left[Y_2\left(G_{m_1|a_1^m a_1^y}^{r_0'} = m_1\right) \middle| R_0 = r_0, a_1^y\right] P(a_1^y)$

$= \sum_{m_1, a_1^y, a_1^m, n_1} E[Y_2 | R_0 = r_0, m_1, n_1, a_1^m, a_1^y]$
$\quad \times \{P(M_1 = m_1 | R_0 = r_0, a_1^m, a_1^y) - P(M_1 = m_1 | R_0 = r_0', a_1^m, a_1^y)\}$
$\quad \times P(n_1 | R_0 = r_0, a_1^m, a_1^y) P(a_1^m | R_0 = r_0, a_1^y) P(a_1^y)$ (10)

Note that (10) does not generally equal the disparity reduction (2) minus (1) because the first intervention $G_{m_1|a_1^m a_1^y}^{r_0}$, within levels of the allowables $A_1^y$ and $A_1^m$ breaks any dependence of treatment intensification $M_1$ on non-allowables $N_1$, whereas in the observed scenario this dependence is present. They are equivalent when $P(M_1 = m_1 | R_0 = r_0, a_1^m, a_1^y) = P(M_1 = m_1 | R_0 = r_0, n_1, a_1^m, a_1^y)$.

Suppose we estimate this difference where $A_1^y$, the covariates deemed both outcome- and target-allowable include $X_0^{g,s}$. $A_1^m$ is left empty so that no additional variables are considered target-allowable, and $N_1$ includes all other variables needed to establish conditional exchangeability for $M_1$ (i.e., $L_1, X_0^{e,i,d}$). Now, under these allowability choices, the difference between the first and second interventions is:

$\sum_{a_1^y} E\left[Y_2\left(G_{m_1|x_0^{g,s}}^{r_0} = m_1\right) \middle| R_0 = r_0, x_0^{g,s}\right] P(x_0^{g,s}) - \sum_{a_1^y} E\left[Y_2\left(G_{m_1|x_0^{g,s}}^{r_0'} = m_1\right) \middle| R_0 = r_0, x_0^{g,s}\right] P(x_0^{g,s})$

$= \sum_{m_1, l_1, x_0^{g,s,e,i,d}} E[Y_2 | R_0 = r_0, m_1, l_1, x_0^{g,s,e,i,d}]$
$\quad \times \{P(M_1 = m_1 | R_0 = r_0, x_0^{g,s}) - P(M_1 = m_1 | R_0 = r_0', x_0^{g,s})\}$
$\quad \times P(l_1, x_0^{e,i,d} | R_0 = r_0, x_0^{g,s}) P(x_0^{g,s})$

This last expression is equivalent to the identification formula for the interventional indirect effect of Vansteelandt and Daniel for the terminal mediator when applied to our motivating example. Again, this does not estimate the disparity reduction because $P(M_1 = m_1 | R_0 = r_0, x_0^{g,s}) \neq P(M_1 = m_1 | R_0 = r_0, l_1, x_0^{g,s,e,i,d})$.

Interventional Analogue of the Path-Specific Indirect Effect III

Suppose we estimate the disparity reduction where $A_1^y$, the covariates deemed both outcome- and target-allowable, include $X^{g,s}$. $A_1^m$, the additional covariates deemed target allowable, includes all other covariates (i.e., $L_1, X_0^{e,i,d}$). $N_1$ is left empty since conditional exchangeability among blacks has been established for $M_1$ given $A_1^m$ and $A_1^y$. The disparity reduction is identified by the non-parametric expressions and weighting approaches of Zheng and van der Laan and also Miles et al.

*Non-parametric*

$$\psi^{red} = \sum_{m_1, l_1, x_0^{g,s,e,i,d}} E[Y_2 | R_0 = r_0, m_1, l_0, x_0^{g,s,e,i,d}]$$
$$\times \{P(M_1 = m_1 | R_0 = r_0, m_1, l_1, x_0^{g,s,e,i,d}) - P(M_1 = m_1 | R_0 = r_0', m_1, l_1, x_0^{g,s,e,i,d})\}$$
$$\times P(l_1, x_0^{e,i,d} | R_0 = r_0, x_0^{g,s})$$
$$\times P(x_0^{g,s})$$

A conditional expression is obtained by removing the integration over $x_0^{g,s}$. This is equivalent to the non-parametric expression of a path-specific effect discussed in Jackson 2018.

*Ratio of Mediator Probability Weighting*

$$\psi^{red} = E\big[E[Y_2 \times w_{r_0} | r_0, m_1, l_1, x_0^{g,s,e,i,d}] | r_0\big] - E\big[E[Y_2 \times w_{r_0}^{rmpw} | r_0, m_1, l_1, x_0^{g,s,e,i,d}] | r_0\big]$$

Where

$$w_{r_0} = \frac{P(r_0)}{P(r_0 | x_0^{g,s})} \qquad w_{r_0}^{rmpw} = \frac{P(M_1 = m_{1j} | R_0 = r_0', l_1, x_0^{g,s,e,i,d})}{P(M_1 = m_{1j} | R_0 = r_0, l_1, x_0^{g,s,e,i,d})} \times w_{r_0}$$

A conditional expression is obtained by conditioning the outer expectation on $x_0^{g,s}$ and setting $w_r = w_r' = 1$. This is related to the weighting approach of Zheng and Van der Laan (2017).

*Inverse Odds Ratio Weighting*

$$\psi^{red} = E\big[E[Y_2 \times w_{r_0} | r_0, m_1, l_1, x_0^{g,s,e,i,d}] | r_0\big] - E\big[E[Y_2 \times w_{r_0}^{iorw} | r_0, m_1, l_1, x_0^{g,s,e,i,d}] | r_0\big]$$

Where

$$w_{r_0} = \frac{P(r_0)}{P(r_0 | x_0^{g,s})} \qquad w_{r_0}^{iorw} = \frac{\frac{P(R = r_0' | m_{1j}, l_1, x_0^{g,s,e,i,d})}{P(R = r_0 | m_{1j}, l_1, x_0^{g,s,e,i,d})}}{\frac{P(R = r_0' | l_1, x_0^{g,s,e,i,d})}{P(R = r_0 | l_1, x_0^{g,s,e,i,d})}} \times w_{r_0}$$

A conditional expression is obtained by conditioning the outer expectation on $x_0^{g,s}$ and setting $w_r = w_r' = 1$. This is equivalent to the "m-ratio" weighting approach proposed by Miles et al. albeit under an alternate coding for race $R_0$. (Note that the specification by Miles et al. would code blacks as $R_0 = r_0'$ and whites as $R_0 = r_0$, mapping to a path-specific effect whose analog imagines an intervention upon whites by fixing the conditional distribution of the target to match that of blacks). Our coding scheme maps to an identification formula for a path-specific effect discussed in Jackson 2018, wherein blacks are intervened upon by fixing the conditional distribution of the target to match that of whites.

### "Detailed" Oaxaca-Blinder Decomposition

Suppose we estimate the disparity reduction where no covariates are deemed outcome- or target-allowable. All covariates are included in $N_1$ to establish exchangeability for $M_1$. The disparity reduction is identified by a "detailed" Oaxaca-Blinder Decomposition implemented with linear models.

The non-parametric formula is:

$$\psi^{red} = \sum_{m_1, l_1, x_0^{g,s,e,i,d}} E[Y_2 | R_0 = r_0, m_1, l_0, x_0^{g,s,e,i,d}]$$
$$\times \{P(M_1 = m_1 | R_0 = r_0, l_1, x_0^{g,s,e,i,d}) - P(M_1 = m_1 | R_0 = r_0')\}$$
$$\times P(l_1, x_0^{g,s,e,i,d} | R_0 = r_0)$$

Consider the following linear models:

$$E[Y_2 | R_0 = r_0, m_{1j}, l_1, x_0^{g,s,e,i,d}] = \beta_0^{r_0} + \sum_{j \neq ref} \beta_{1j}^{r_0} I(M_1 = m_{1j}) + \beta_2^{r_0} L_1 + \sum_k \beta_3^k X^k$$

$$E[Y_2 | R_0 = r_0', m_{1j}, l_1, x_0^{g,s,e,i,d}] = \beta_0^{r_0'} + \sum_{j \neq ref} \beta_{1j}^{r_0'} I(M_1 = m_{1j}) + \beta_2^{r_0'} L_1 + \sum_k \beta_3^k X^k$$

where, with a slight abuse of notation, $X^k$ is the kth element of $x_0^{g,s,e,i,d}$.

It follows from the arguments of Jackson and VanderWeele 2018 we that:

$$\psi^{red} = \sum_{j \neq ref} \beta_{1j}^{r_0} \{P(M = m_{1j} | R_0 = r_0) - P(M = m_{1j} | R_0 = r_0')\}$$

This is the typical formulation of a detailed Oaxaca-Blinder Decomposition under linear models. Alternate implementations of the Oaxaca-Blinder Decomposition make different allowability choices. For example, suppose we estimate the disparity reduction where no covariates are deemed outcome-allowable but all covariates are considered target-allowable, leaving $N_1$ empty. The disparity reduction is identified by the following non-parametric formula which leads to adaptations of the weighting estimators of Dinardo et al. (a form of ratio of mediator probability weighting) and Barsky et al. (a form of inverse odds ratio weighting).

*Non-parametric*

$$\psi^{red} = \sum_{m_1, l_1, x_0^{g,s,e,i,d}} E[Y_2 | R_0 = r_0, m_1, l_0, x_0^{g,s,e,i,d}]$$
$$\times \{P(M_1 = m_1 | R_0 = r_0, l_1, x_0^{g,s,e,i,d}) - P(M_1 = m_1 | R_0 = r_0', l_1, x_0^{g,s,e,i,d})\}$$
$$\times P(l_1, x_0^{g,s,e,i,d} | R_0 = r_0)$$

*Ratio of Mediator Probability Weighting*

$$\psi^{red} = E\left[E[Y_2 \times w_{r_0} | r_0, m_1, l_1, x_0^{g,s,e,i,d}] | r_0\right] - E\left[E[Y_2 \times w_{r_0}^{rmpw} | r_0, m_1, l_1, x_0^{g,s,e,i,d}] | r_0\right]$$

Where

$$w_0^r = 1 \qquad w_{r_0}^{rmpw} = \frac{P(M_1 = m_{1j} | R_0 = r_0', l_1, x_0^{g,s,e,i,d})}{P(M_1 = m_{1j} | R_0 = r_0, l_1, x_0^{g,s,e,i,d})} \times w_{r_0}$$

This is equivalent to an extension of the weighting approach proposed by Dinardo, Fortin and Lemieux where the conditioning events of the numerator and denominator of $w_{r_0}^{rmpw}$ include all covariates.

*Inverse Odds Ratio Weighting*

$$\psi^{red} = E\big[E[Y_2 \times w_{r_0}|r_0, m_1, l_1, x_0^{g,s,e,i,d}]|r_0\big] - E\big[E[Y_2 \times w_{r_0}^{iorw}|r_0, m_1, l_1, x_0^{g,s,e,i,d}]|r_0\big]$$

Where

$$w_0^r = 1 \qquad w_{r_0}^{iorw} = \frac{\dfrac{P(R = r_0'|m_{1j}, l_1, x_0^{g,s,e,i,d})}{P(R = r_0|m_{1j}, l_1, x_0^{g,s,e,i,d})}}{\dfrac{P(R = r_0'|l_1, x_0^{g,s,e,i,d})}{P(R = r_0|l_1, x_0^{g,s,e,i,d})}} \times w_{r_0}$$

This is equivalent to an extension of the weighting approach proposed by Barsky et al., and also one discussed by Dinardo, Fortin and Lemieux, where the conditioning events of the numerator and denominator of $w_{r_0}^{iorw}$ include all covariates.